\documentclass[11pt]{article}
\usepackage{amssymb}
\usepackage{epsfig}
\newlength{\bredde}
\def\slash#1{\settowidth{\bredde}{$#1$}\ifmmode\,\raisebox{.15ex}{/}
\hspace*{-\bredde} #1\else$\,\raisebox{.15ex}{/}\hspace*{-\bredde} #1$\fi}
\newcommand{\be}{\begin{equation}}
\newcommand{\ee}{\end{equation}}
\newcommand{\bea}{\begin{eqnarray}}
\newcommand{\eea}{\end{eqnarray}}
\newcommand{\nn}{\nonumber}
\newcommand{\sect}[1]{\setcounter{equation}{0}\section{#1}}

\newcommand{\wlm}{w^{[\ell,m]}}

\newcommand{\wm}{w^{[0,m]}}

\newcommand{\pilm}{\pi^{[\ell,m]}}

\newcommand{\qm}{q^{[0,m]}}
\newcommand{\qlm}{q^{[\ell,m]}}

\begin{document}
\topmargin -1.4cm
\oddsidemargin -0.8cm
\evensidemargin -0.8cm
\title{\Large{{\bf 
Ratios of characteristic polynomials in complex matrix models 
}}}

\vspace{2.5cm}

\author{{\sc G. Akemann}$^*$ and {\sc A. Pottier}$^\dag$\\~\\~\\
$^*$Service de Physique Th\'eorique, CEA/DSM/SPhT Saclay\\
Unit\'e de recherche associ\'ee CNRS/SPM/URA 2306\\
F-91191 Gif-sur-Yvette Cedex, France
\\~\\
$^\dag$Laboratoire de Physique Th\'eorique, Ecole Normale Sup\'erieure\\
24 rue Lhomond, F-75231 Paris Cedex 05, France
}

\date{}
\maketitle
\vfill
\begin{abstract}
We compute correlation functions of inverse powers and ratios of 
characteristic polynomials for random matrix models with complex eigenvalues.
Compact expressions are given in terms of orthogonal polynomials in the 
complex plane as well as their Cauchy transforms, generalizing previous 
expressions for real eigenvalues. 
We restrict ourselves to ratios of characteristic 
polynomials over their complex conjugate.
\end{abstract}

\vfill

\begin{flushleft}
SPhT T04/047\\
\end{flushleft}
\thispagestyle{empty}
\newpage

\renewcommand{\thefootnote}{\arabic{footnote}}
\setcounter{footnote}{0}

\sect{Main Results}\label{intro}

The theory of random matrices has found many applications in different 
branches of physics  \cite{GMW} as well as in mathematics. One possibility to 
study the correlation functions of matrix eigenvalues, which can then be mapped
to various physical or mathematical quantities, is to compute ratios 
of characteristic polynomials as their generating functional. 
Such correlation functions can also be studied in their own right, as they
enjoy a direct physical interpretation as well.
While the most general generating functional is known for real eigenvalues 
much less was known until recently for complex eigenvalues. Characteristic 
polynomials of the corresponding non-hermitian operators play an important 
role for example in scattering in Quantum Chaos, as reviewed 
in \cite{FS}, or in Quantum Chromodynamics \cite{HJV,AFV}.  

Our purpose is to generalize the result \cite{AV} for arbitrary products 
of characteristic polynomials and their complex 
conjugates of not necessarily the same order, to ratios of such objects. 
We restrict ourselves to the case of 
matrix models with a complex eigenvalue representation, to make 
the technique of orthogonal polynomials available. 
Its difficulty in the complex plane is that in 
general neither the three-step recursion relation nor the Christoffel-Darboux
formula hold in general. We will still be able to show that
part of the results of \cite{BDS} generalize to complex eigenvalues, 
where streamline proofs of the previous achievements \cite{BH,FS2} for real 
eigenvalues are given.

The partition function of a complex (matrix) eigenvalue model is defined as
\be
Z_N \ \equiv\ \prod_{i=1}^N \left( \int_D dw(z_i,\bar{z}_i)\right)
\left|\Delta_N(\{z\})\right|^2\ \ \ ,\ 
\Delta_N(\{z\})\ \equiv\ \prod_{i>j}^N(z_i-z_j)\ ,
\label{Zev}
\ee
where we have introduced the Vandermonde determinant $\Delta_N(\{z\})$ 
stemming from the Jacobian of the diagonalization.
We suppose that the probability measure 
$dw(z,\bar{z})=dzd\bar{z}\, w(z,\bar{z})$ 
can be written in terms of eigenvalues and factorizes, and thus that 
eq. (\ref{Zev})
is invariant under permutations of the eigenvalues $z_{i=1,\ldots,N}$.
The weight function $w(z,\bar{z})$ depending both on $z$ and $\bar{z}$
shall be strictly positive on the domain of 
integration $D$. $D$ can be either a compact set or the full complex plane. 
Furthermore we require that all moments exist, $\int_D dw(z,\bar{z}) z^k
<\infty$ for all $k=0,1,\ldots$ .
Examples for such general weight functions and domains $D$ are given in 
\cite{AV}. 
Under such conditions a unique set of orthogonal polynomials in the complex 
plane can be introduced using the Gram-Schmidt procedure. 
The monic polynomials $\pi_k(z)=z^k+\ldots$ of degree $k$ 
follow from
\be
\int_D dw(z,\bar{z})\ \pi_k(z) \overline{\pi_j(z)}\ =\ \delta_{k,j}\, r_k\ ,
\label{OP}
\ee
with their squared norms $r_k>0$ being strictly positive. 
We note that in contrast to the weight the polynomials $\pi_k(z)$ only depend 
on $z$ and not its complex conjugate $\bar{z}$.
The corresponding Cauchy transform in the complex plane is given by 
\be
h_n(\bar{\epsilon})\equiv \frac{1}{2\pi i}
\int_D dw(z,\bar{z})\ \frac{\pi_n(z)}{\bar{z}-\bar{\epsilon}}\ .
\label{Cauchy}
\ee
Note that the pole $\bar{\epsilon}$ is an integrable singularity in the 
complex plane.
If we want to allow for the hermitian limit to be taken we have to require
$\bar{\epsilon}\notin \bar{D}$. In that case $D$ has to become compact 
at least in the large-$N$ limit.
The Cauchy transforms 
$h_n(\bar{\epsilon})$ only depend on $\bar{\epsilon}$ and not on $\epsilon$.
Another possible Cauchy transform, dividing  $\pi_n(z)$ by $z-\epsilon$, is 
not needed as it can be expressed through the $\pi_{k<n}(z)$ and 
$h_0(\epsilon)$.
Expectation values of observables of symmetric functions of the eigenvalues 
$f_N\equiv f_N(z_1,\ldots,z_N)$ can be defined as 
\be
\langle f_N \rangle_w\ \equiv\ \frac{1}{Z_N}\prod_{i=1}^N 
\left( \int_D dw(z_i,\bar{z}_i)\right)
\ f_N(z_1,\ldots,z_N)\ \left|\Delta_N(\{z\})\right|^2\ .
\label{vev}
\ee
Our objects of interest are ratios of
characteristic polynomials $D_N[\mu]$ and their conjugates,
\be
D_N[\mu] \ \equiv\ \prod_{i=1}^N(\mu-z_i)\ \ \mbox{and}\ \ 
D_N^\dag[\bar{\epsilon}]\ \equiv\ \prod_{i=1}^N(\bar{\epsilon}-\bar{z}_i)\ .
\label{chP}
\ee
We can now state our main result, generalizing \cite{FS2} (see also 
theorem 2.13 of \cite{BDS}).

\indent

{\bf Theorem:} Let  $\{\mu_j\}_{j=1,\ldots,L}$ and 
$\{\bar{\epsilon}_k\}_{k=1,\ldots,M}$ be pairwise 
non-degenerate complex variables. For $0\le M\le N$ it holds:
\begin{equation}
\left\langle 
\frac{\prod_{j=1}^L D_N[\mu_j]}{\prod_{k=1}^MD_N^{\dag}[\bar{\epsilon}_k]}
\right\rangle
_{w}=
\frac{(-1)^{\frac{M(M-1)}{2}}\prod\limits_{j=N-M}^{N-1}
\left(\frac{2\pi }{i\,r_j}\right)
}{\Delta_L(\{\mu\})\Delta_M(\{\bar{\epsilon}\})}
\left|\begin{array}{ccc}
  h_{N-M}(\bar{\epsilon}_1) & \ldots & h_{N+L-1}(\bar{\epsilon}_1) \\
  \vdots &  &  \\
  h_{N-M}(\bar{\epsilon}_M) & \ldots & h_{N+L-1}(\bar{\epsilon}_M)  \\
  \pi_{N-M}(\mu_1) & \ldots & \pi_{N+L-1}(\mu_1) \\
  \vdots &  &  \\
  \pi_{N-M}(\mu_L) & \ldots & \pi_{N+L-1}(\mu_L)
\end{array}\right| .
\label{theorem}
\end{equation}

The following two special cases are worth to be mentioned. For $L=0$ and 
$M\neq0$, that is 
for inverse powers only, we obtain a determinant composed purely of 
Cauchy transforms eq. (\ref{Cauchy}), generalizing the results of \cite{FS2}
(see also theorem 2.10 in \cite{BDS}). In the opposite case, for $L\neq0$ and 
$M=0$ with only products, we partially recover the result of \cite{AV} in 
terms of polynomials only. 
The theorem as well as the special cases trivially carry over to the complex 
conjugate expressions. The limit of coinciding variables, e.g. $\mu_i=\mu_j$,
can be easily taken, leading to derivatives of the polynomials and 
Cauchy transforms.


\sect{Proofs}\label{result}

The proof will very closely follow the steps taken in \cite{BDS}.
Due to the Heine-formula it is well known that orthogonal polynomials 
with respect to a given weight can be expressed through characteristic 
polynomials,
\be
\left\langle  D_N[\mu] \right \rangle_w\ =\  \pi_N(\mu)\ .
\label{Heine}
\ee
In the following it will be useful to consider
a generalized measure,
\be
 d\wlm(z,\bar{z}) \ \equiv\ 
  \frac{\prod_{j=1}^\ell (\mu_j-z)}{\prod_{k=1}^m(\bar{\epsilon}_k-\bar{z})}
  dw(z,\bar{z})\ , \qquad
  \ell,m\ge 0 \ ,
\label{genweight}
\ee
as well as the corresponding quantities eqs. (\ref{Zev}) --
(\ref{vev})\footnote{The superscript $[0,0]$ is sometimes dropped, 
corresponding to the previous definitions.}.
The expectation value in the theorem is then proportional to 
``orthogonal polynomials'' with respect to eq. (\ref{genweight}),
$\pi_N^{[L-1,M]}(\mu_L)$. We will explicitly construct such polynomials 
by requiring
\be 
\int_D  d\wlm(z,\bar{z})\ \pilm_j(z)\ \bar{z}^k\ =\ 0\ ,
  \qquad j>k\ge 0\ .
\label{biOP}
\ee
They can be interpreted as bi-orthogonal polynomials \cite{B1}.
Let us stress however, that our result eq. (\ref{theorem}) can be entirely 
formulated in terms of truly orthogonal polynomials and their 
Cauchy transforms, which form a bona fide scalar product. 

Our proof goes in four steps. We consecutively construct the polynomials 
$\pi_n^{[\ell,0]}(z)$, $\pi_n^{[0,m]}(z)$, the Cauchy transform of the latter
$h_n^{[0,m]}(z)$, and finally $\pi_n^{[\ell,m]}(z)$. In this way we first show 
the special cases $M=0$ and $L=0$ respectively, before arriving 
at eq. (\ref{theorem}).

\indent

{\bf Step 1.} Let us define for $\ell\geq 1$
\be
q_n^{[\ell,0]}(z)\equiv  \left|\begin{array}{ccc}
  \pi_n(\mu_1) & \cdots & \pi_{n+\ell}(\mu_1) \\
  \vdots &  &  \\
  \pi_n(\mu_\ell) & \ldots & \pi_{n+\ell}(\mu_\ell) \\
  \pi_n(z) & \ldots & \pi_{n+\ell}(z)
\end{array}\right| ,
\label{q1}
\ee
for which it holds
\be
\int_D dw(z,\bar{z})\  q_n^{[\ell,0]}(z)\ \bar{z}^j \ =\ 0\ , \ \ \  0
\le j \le n-1\ .
\ee
Because of  $q_n^{[\ell,0]}(\mu_j)=0, \; j=1,\cdots,\ell$, the ratio
$\frac{q_n^{[\ell,0]}(z)}{(\mu_1-z)\cdots(\mu_\ell-z)}$ 
is a polynomial of degree\footnote{The fact that the degree is $n$ and 
not less can be shown by induction.} $n$. It can thus be written as a linear 
combination of the polynomials $\pi_{0,\ldots,n}(z)$ forming a complete set. 
Consequently 
\be
\int_D  dw^{[\ell,0]}(z,\bar{z}) \left[\frac{q_n^{[\ell,0]}(z)}{(\mu_1-z)\ldots
(\mu_\ell-z)}\right] \bar{z}^j=0,\;\;\mbox{for}\ 0\leq j<n\ .
\ee
In order to achieve a monic normalization we can expand eq. (\ref{q1}) with 
respect to the last row, and take $z\to\infty$ to read off
the generalized Christoffel formula in the complex plane
\begin{equation}
\pi_{n}^{[\ell,0]}(z)=\frac{1}{(z-\mu_1)\ldots (z-\mu_\ell)}\;
\left|\begin{array}{ccc}
  \pi_n(\mu_1) & \cdots & \pi_{n+\ell}(\mu_1) \\
  \vdots &  &  \\
  \pi_n(\mu_\ell) & \ldots & \pi_{n+\ell}(\mu_\ell) \\
  \pi_n(z) & \ldots & \pi_{n+\ell}(z)
\end{array}\right| \cdot
\left|\begin{array}{ccc}
  \pi_{n}(\mu_1) & \ldots & \pi_{n+\ell-1}(\mu_1) \\
  \vdots &  &  \\
  \pi_{n}(\mu_\ell) & \ldots & \pi_{n+\ell-1}(\mu_\ell)
\end{array}\right|^{-1}
\ .
\label{Christoffel1}
\end{equation}
These polynomials were previously computed in \cite{B1}.
The denominator is non-vanishing due to the non-degeneracy of the $\mu_j$.
Due to the relation eq. (\ref{Heine}) for the general weight 
$\pi_N^{[j,0]}(\mu_{j+1})=\left\langle
D_N[\mu_{j+1}]\right\rangle_{w^{[j,0]}}$,
and the observation that the expectation value can be written as a telescope 
product, 
\begin{equation}
\left\langle\prod\limits_{j=1}^{L}D_N[\mu_j]\right\rangle_{w}=
\prod\limits_{j=1}^L\left\langle
D_N[\mu_j]\right\rangle_{w^{[j-1,0]}}=
\prod\limits_{j=0}^{L-1}\pi_{N}^{[j,0]}(\mu_{j+1})\ ,
\label{telescope1}
\end{equation}
we can deduce eq. (\ref{theorem}) for $M=0$ upon using eq. 
(\ref{Christoffel1}).

\indent

{\bf Step 2.} Next we define
\be
q_n^{[0,m]}(z)\equiv \left|\begin{array}{ccc}
  h_{n-m}(\bar{\epsilon}_1) & \ldots & h_n(\bar{\epsilon}_1) \\
  \vdots &  &  \\
  h_{n-m}(\bar{\epsilon}_m) & \ldots & h_n(\bar{\epsilon}_m) \\
  \pi_{n-m}(z) & \ldots & \pi_n(z)
\end{array}\right|,
\label{q2}
\ee
which automatically implies 
\begin{equation} 
\int_D dw(z,\bar{z})
\frac{\qm_n(z)}{\bar{z}-\bar{\epsilon}_j}\ =\ 0\ , \qquad j=1,
\cdots,m\ .
\label{q2orth}
\end{equation}
For $0\le j<n$ we can decompose 
\be
  \frac{\bar{z}^j}{\prod_{k=1}^m(\bar{\epsilon}_k-\bar{z})} = \sum_{k=1}^m
  \frac{a_k}{\bar{\epsilon}_k-\bar{z}} + p(\bar{z})\ ,
\ee
where $p(\bar{z})$ is a polynomial of degree $< n - m$. 
Consequently 
\be 
\int_D  d\wm(z,\bar{z}) \ \qm_n(z)\  \bar{z}^j=  \sum_{k=1}^m a_k
  \int_D  dw(z,\bar{z}) \frac{\qm_n(z)}{\bar{\epsilon}_k-\bar{z}} + 
\int_D dw(z,\bar{z})\  \qm_n(z)\  p(\bar{z}) =0
\ee
vanishes due to eq. (\ref{q2orth}) in the first sum and orthogonality in the 
second term. In monic normalization we thus have the generalized Uvarov formula
\begin{equation}
\pi_{n}^{[0,m]}(z)= 
\left|\begin{array}{ccc}
  h_{n-m}(\bar{\epsilon}_1) & \ldots & h_n(\bar{\epsilon}_1) \\
  \vdots &  &  \\
  h_{n-m}(\bar{\epsilon}_m) & \ldots & h_n(\bar{\epsilon}_m) \\
  \pi_{n-m}(z) & \ldots & \pi_n(z)
\end{array}\right|\cdot
\left|\begin{array}{ccc}
  h_{n-m}(\bar{\epsilon}_1) & \ldots & h_{n-1}(\bar{\epsilon}_1) \\
  \vdots &  &  \\
 h_{n-m}(\bar{\epsilon}_m) & \ldots & h_{n-1}(\bar{\epsilon}_m)
\end{array}\right|^{-1}.
\label{Uvarov}
\end{equation}

\indent

{\bf Step 3.} Let  $0\leq m\leq n$. 
The Cauchy transform of eq. (\ref{Uvarov}) for the measure $d\wm(z)$ can be 
expressed in terms of the $h_n(\bar{z})$ by writing 
\be
h_n^{[0,m]}(\bar{\epsilon})=\frac{1}{2\pi i}
\int_D d\wm(z,\bar{z}) \frac{\pi_n^{[0,m]}(z)}{\bar{z}-\bar{\epsilon}}
=\sum_{j=1}^{m+1}\frac1{2\pi i}
\prod_{k\neq j}\frac{(-1)^m}{\bar{\epsilon}_j-\bar{\epsilon}_k}
\int_D dw(z,\bar{z}) \frac{\pi_n^{[0,m]}(z)}{\bar{z}-\bar{\epsilon}_j}\ .
\ee
Only the term in $\bar{\epsilon}
\equiv \bar{\epsilon}_{m+1}$ is non-vanishing, and thus we 
obtain from  eq. (\ref{Uvarov})
\begin{equation}
h_n^{[0,m]}(\bar{\epsilon})=\frac{(-1)^m}{(\bar{\epsilon}-\bar{\epsilon}_m)
\ldots
(\bar{\epsilon}-\bar{\epsilon}_1)}\;
\left|\begin{array}{ccc}
  h_{n-m}(\bar{\epsilon}_1) & \ldots & h_n(\bar{\epsilon}_1) \\
  \vdots &  &  \\
  h_{n-m}(\bar{\epsilon}_m) & \ldots & h_n(\bar{\epsilon}_m) \\
  h_{n-m}(\bar{\epsilon}) & \ldots & h_n(\bar{\epsilon})
\end{array}\right|
\cdot
\left|\begin{array}{ccc}
  h_{n-m}(\bar{\epsilon}_1) & \ldots & h_{n-1}(\bar{\epsilon}_1) \\
  \vdots &  &  \\
  h_{n-m}(\bar{\epsilon}_m) & \ldots & h_{n-1}(\bar{\epsilon}_m)
\end{array}\right|^{-1} .
\label{Christoffel2}
\end{equation}
These expressions can be used in the following identity for an inverse 
characteristic polynomial
\bea
\left\langle
 D_N^{\dag}[\bar{\epsilon}]^{-1}\right\rangle_w
&=&
\frac1{Z_N} \prod_{i=1}^N\left(\int_D dw(z_i,\bar{z}_i)\right)\sum_{j=1}^{N}
\frac{|\Delta_N(\{z\})|^2}{\prod\limits_{k\neq
j}\bar{z}_j-\bar{z}_k}\;\frac{1}{\bar{\epsilon}-\bar{z}_j}
= N\frac{Z_{N-1}}{Z_N}\int_D dw(z_N,\bar{z}_N)
\frac{\pi_{N-1}(z_N)}{\bar{\epsilon}-\bar{z}_N}\nn\\
&=& -2\pi i N\frac{Z_{N-1}}{Z_N}\ h_{N-1}(\bar{\epsilon})\ .
\label{Heine-1}
\eea
After decomposing the inverse product we have used the identity 
$\frac{|\Delta_N(\{z\})|^2}{\prod\limits_{k<N}\bar{z}_N-\bar{z}_k}=|
\Delta_{N-1}(\{z\})|^2\!\!\prod\limits_{k<N}\!(z_N-z_k)$ 
as well as the permutation 
symmetry of the integrand to deduce this generalized Heine formula for the 
Cauchy transform. In order to apply eq. 
(\ref{Christoffel2}) we rewrite identically
\begin{equation}
\left\langle\prod\limits_{j=1}^{M}D_N^{\dag}
\left[\bar{\epsilon}_j\right]^{-1}\right\rangle_w
=\frac{Z_N^{[0,M]}}{Z_{N-1}^{[0,M-1]}}\;
\frac{Z_{N-1}^{[0,M-1]}}{Z_{N-2}^{[0,M-2]}}\ldots
\frac{Z_{N-M}^{[0,0]}}{Z_{N}^{[0,0]}}\ .
\label{telescope2}
\end{equation}
From eq. (\ref{Heine-1}) valid for the general weight eq. (\ref{genweight})
we can conclude 
\be
\frac{Z_{N-k}^{[0,m]}}{Z_{N-k-1}^{[0,m-1]}}= -2\pi i
(N-k)\;h_{N-k-1}^{[0,m-1]}(\bar{\epsilon}_{m}) \ ,
\label{Z-h}
\ee
writing the expectation value as a ratio of partition functions. 
It is a well known fact that the partition function eq. (\ref{Zev}) can 
be expressed in terms of the norms eq. (\ref{OP}), 
$Z_N=N!\prod_{j=0}^{N-1}r_j$, or equivalently 
$(N-k)=Z_{N-k}/(r_{N-k}Z_{N-k-1})$. Replacing this factor in 
eq. (\ref{Z-h}) and inserting it into eq. (\ref{telescope2}) all partition 
functions cancel and we obtain
\begin{equation}
\left\langle\prod\limits_{j=1}^{M}D_N^{\dag}
\left[\bar{\epsilon}_j\right]^{-1}\right\rangle_w
=\prod\limits_{j=1}^{M}\frac{-2\pi i}{r_{N-j}}\ 
h_{N-j}^{[0,M-j]}(\bar{\epsilon}_{M-j+1})\ .
\label{step3}
\end{equation}
Together with eq. (\ref{Christoffel2})
this leads to the theorem eq. (\ref{theorem}) in the special case of $L=0$.

\indent

{\bf Step 4.} We can now give the polynomials with respect to the most 
general weight eq. (\ref{genweight}), 
\begin{equation}
\pi_{n}^{[\ell,m]}(z)=\frac{1}{(z-\mu_\ell)\ldots (z-\mu_1)}\;
\left|\begin{array}{ccc}
  h_{n-m}(\bar{\epsilon}_1) & \ldots & h_{n+\ell}(\bar{\epsilon}_1) \\
  \vdots &  &  \\
  h_{n-m}(\bar{\epsilon}_m) & \ldots & h_{n+\ell}(\bar{\epsilon}_m)  \\
  \pi_{n-m}(\mu_1) & \ldots & \pi_{n+\ell}(\mu_1) \\
  \vdots &  &  \\
  \pi_{n-m}(\mu_\ell) & \ldots & \pi_{n+\ell}(\mu_\ell)\\
  \pi_{n-m}(z) & \ldots & \pi_{n+\ell}(z)
\end{array}\right|
\cdot
\left|\begin{array}{ccc}
  h_{n-m}(\bar{\epsilon}_1) & \ldots & h_{n+\ell-1}(\bar{\epsilon}_1) \\
  \vdots &  &  \\
 h_{n-m}(\bar{\epsilon}_m) & \ldots & h_{n+\ell-1}(\bar{\epsilon}_m)  \\
  \pi_{n-m}(\mu_1) & \ldots & \pi_{n+\ell-1}(\mu_1) \\
  \vdots &  &  \\
  \pi_{n-m}(\mu_\ell) & \ldots & \pi_{n+\ell-1}(\mu_\ell)
\end{array}\right|^{-1}\!,
\label{Christoffel3}
\end{equation}
holding for $0\le m\le n$. 
If we define by $\qlm_n(z)$ the determinant in the numerator it holds 
\bea
0&=& q_n^{[\ell,m]}(\mu_1)=\ldots=q_n^{[\ell,m]}(\mu_\ell)\ ,\nn\\
0&=&\int_D dw(z,\bar{z})\frac{q_{n}^{[\ell,m]}(z)}{\bar{\epsilon}_1-\bar{z}}
=\ldots
=\int_D dw(z,\bar{z})\frac{q_{n}^{[\ell,m]}(z)}{\bar{\epsilon}_m-\bar{z}}\ .
\eea
This can be seen following the same lines as in the previous steps, and thus 
that eq. (\ref{Christoffel3}) is correct including its normalization.
In order to prove eq. (\ref{theorem}) we decompose 
\be
\left\langle\frac{\prod_{j=1}^L D_N[\mu_j]}{\prod_{k=1}^M D_N^{\dag}
[\bar{\epsilon}_k]}
\right\rangle_{w}= \left\langle\prod\limits_{j=1}^L D_N[\mu_j]
\right\rangle_{w^{[0,M]}}\cdot
\left\langle\prod\limits_{j=1}^M D_N^{\dag}
[\bar{\epsilon}_j]^{-1} \right\rangle_{w} 
= 
\prod\limits_{j=0}^{L-1}\pi_{N}^{[j,M]}(\mu_{j+1})
\left\langle\prod\limits_{j=1}^MD_N^{\dag}[\bar{\epsilon}_j]^{-1}
\right\rangle_{w}
\ . 
\ee
Inserting eq. (\ref{Christoffel3}) and the previous result eq. (\ref{step3})
from step 3 the theorem eq. (\ref{theorem}) follows.

\sect{Conclusions}\label{conclusions}   

We have computed the correlation functions of arbitrary products 
of characteristic polynomials over arbitrary products 
of complex conjugate characteristic polynomials for random matrix models 
with complex eigenvalues. This extends previous results for 
only products of mixed  characteristic polynomials and their conjugates 
\cite{AV} (see also \cite{B1}). From the result \cite{AV} we expect that 
more general correlation functions of mixed ratios will contain both 
polynomials and Cauchy transforms as well as the various kernels constructed 
out of them, as introduced in \cite{FS2} in the real case. 
This would be needed to compare for example to the matrix model result 
(without eigenvalue representation) \cite{SV}, where the average of  the 
inverse of a single characteristic polynomial and its 
complex conjugate was computed.
We hope that our results will help to further clarify the issue of complex 
matrix model universality \cite{A02II}, with 
an extension of the results \cite{FS2,AF} being desirable. 

When writing up our results the preprint \cite{B2} appeared, which 
partly overlaps. There, ratios of mixed characteristic polynomials 
and their conjugates 
are expressed in terms of three different kernels, providing formulas 
of the two-point functions type.

\indent

\noindent
\underline{Acknowledgments}: 
M. Berg\`ere, J.-L. Cornou and G. Vernizzi 
are thanked for useful conversations. The work of G.A. is supported by 
a Heisenberg fellowship of the DFG.


\end{document}